\documentstyle[PASJadd]{PASJ95}
\begin{document}
\title{Subaru Deep Survey I.  Near-Infrared Observations}
\author{
Toshinori {\sc Maihara}\\
{\it Department of Astronomy, Kyoto University, Kitashirakawa, 
Kyoto 606-8502}\\
{\it E-mail(TM): maihara@kusastro.kyoto-u.ac.jp}\\
Fumihide {\sc Iwamuro},
Hirohisa {\sc Tanabe},
Tomoyuki {\sc Taguchi},
Ryuji {\sc Hata}, \\
{\it Department of Physics, Kyoto University, Kitashirakawa, Kyoto 606-8502}\\
Shin {\sc Oya},\\
{\it Communications Research Laboratory, Koganei, Tokyo 184-8975}\\
Nobunari {\sc Kashikawa},
Masanori {\sc Iye},
Satoshi {\sc Miyazaki},
Hiroshi {\sc Karoji}, \\
Michitoshi {\sc Yoshida}\\
{\it Optical and Infrared Astr. Div., National Astronomical 
Observatory, Mitaka, Tokyo 181-8588}\\
Tomonori {\sc Totani}\\
{\it Theoretical Astrophysics Div., National Astronomical
Observatory, Mitaka, Tokyo 181-8588}\\
Yuzuru {\sc Yoshii}\\
{\it Institute of Astronomy, School of Science, University of Tokyo, 
Mitaka, Tokyo 113-0033}\\
Sadanori {\sc Okamura}, 
Kazuhiro {\sc Shimasaku},
Yoshihiko {\sc Saito}\\
{\it Department of Astronomy, University of Tokyo, Hongo, Tokyo 113-0033}\\
Hiroyasu {\sc Ando},
Miwa {\sc Goto}, 
Masahiko {\sc Hayashi},
Norio {\sc Kaifu},
Naoto {\sc Kobayashi}, \\
George {\sc Kosugi},
Kentaro {\sc Motohara},
Tetsuo {\sc Nishimura},
Jun'ichi {\sc Noumaru},\\
Ryusuke {\sc Ogasawara}, 
Toshiyuki {\sc Sasaki}, 
Kazuhiro {\sc Sekiguchi},
Tadafumi {\sc Takata}, \\
Hiroshi {\sc Terada}, 
Takuya {\sc Yamashita}, 
Tomonori {\sc Usuda} \\
{\it Subaru Telescope, National Astron. Observatory, 
650 N. Aohoku Place, Hilo, HI 96720, USA}\\
and \\
Alan T. {\sc Tokunaga} \\
{\it Institute for Astronomy, University of Hawaii, 2680 Woodlawn Dr., 
Honolulu, HI 96822, USA}
}

\abst{
Deep near-infrared images of a blank 2$^{\prime}$ $\times$ 
2$^{\prime}$ section of sky near the Galactic north pole taken by 
Subaru Telescope are presented.  The total integration times of the 
$J$ and $K^{\prime}$ bands are 12.1 hours and 9.7 hours, resulting 
in 5-sigma limiting magnitudes of 25.1 and 23.5 mag, respectively.  
The numbers of sources within these limiting magnitudes 
found with an automated detection procedure are 385 in the $J$ band 
and 350 in $K^{\prime}$.  Based on photometric measurements of these 
sources, we present number count vs.  magnitude relations, 
color vs.  magnitude diagrams, size vs.  color relationships, etc.  
The slope of the galaxy 
number count plotted against the AB magnitude scale is about 0.23 in 
the 22 to 26 AB magnitude range of both bands.  The spatial number 
density of galaxies as well as the slopes in the faint-end region given 
by the Subaru Deep Field (SDF) survey is consistent with those given 
by HST-NICMOS surveys as expressed on the AB magnitude diagram.  Several 
sources having very large $J-K^{\prime}$ color are found including a 
few $K^{\prime}$ objects without detection at $J$.  In addition, a 
number of faint Galactic stars are also detected, most of which are 
assigned to M-subdwarfs, together with a few brown dwarf candidates.  
}

\kword{ cosmology: observations --- cosmology: early universe ---
infrared: galaxies --- infrared: stars --- galaxies: evolution ---
stars: low-mass --- stars: brown dwarfs
}

\maketitle
\thispagestyle{headings}

\section
{Introduction}

Extremely deep imaging of blank fields is a vital method of 
delineating the nature of the early universe and gaining general 
knowledge about the physical conditions at such an early epoch.  In 
this context, the purpose of deep surveys are not only to search for 
bright, peculiar objects at very high redshift, but to also
learn about the overall nature of the early universe.  The 
optical Hubble Deep Field (HDF) images taken by Hubble Space Telescope 
have presented views different from the present day universe.  They 
showed that faint, irregular, and smaller galaxy populations seem to be 
much more abundant, perhaps inherent to the early universe.
However, since the faint, high-redshift objects seen in the optical HDF 
images are deemed to represent rest frame UV emissions, the information 
is predominantly related to UV-luminous sites, 
presumably associated with current star formation, rather than the 
fundamental structure of stellar components in galaxies.  On the other 
hand, a near-infrared deep survey is expected to convey information 
about the basic galactic structure, or in other words, information 
related to the fundamental mass distribution.
Near-infrared observations may also be crucial to probe galaxies in 
the most distant region, because the effect of intergalactic 
reddening, if it occurs, gets smaller at longer wavelengths.  \\

There are already a number of near-infrared surveys (Gardner et al, 1993, 
Bershady et al.  1998, Yan et al.  1998, Thompson et al.  1999), with 
spatial coverage, wavelength bands, and limiting magnitudes differing 
from survey to survey.  The observed wavelengths in the near-infrared 
may correspond to rest frame wavelengths within a much broader 
spectral span, if objects with redshifts of 5 or even larger are 
included.  As for the deepest near-infrared survey, NICMOS images of a 
part of the HDF region have provided deep source counts in both 
the $J$ and $H$ bands (Thompson et al.  1999).  The claimed limiting 
magnitudes are between 27.5 to 28 AB magnitudes at the 80\% 
completeness level.  It is interesting to note that the source 
count vs.  magnitude diagrams of both these bands show an appreciably 
lower number density of galaxies than previous ground-based results.  
Since the sensitivity of the NICMOS imager in the $K$ or $K'$ band is 
limited by the thermal radiation of the telescope, extremely deep 
surveys using ground-based telescopes are important especially in the 
$K'$ band.  Here we present the $K'$ band image as well as that at 
$J$, both currently the deepest images taken by a ground-based 
telescope.

In this report, we concentrate on the details of observations using 
the newly commissioned 8.2 m Subaru Telescope atop Mauna Kea and on 
data analysis, and then present number count diagrams of galaxies in 
two near-infrared bands, $J$ (1.25 $\mu$m) and $K^{\prime}$ (2.13 
$\mu$m).  We also show the detection and identification of stellar 
components found in the SDF survey region.  Cosmological constants are 
assumed to be $H_0=65$ km s$^{-1}$ Mpc$^{-1}$ and $q_0=0.1$ throughout 
this paper.

\section
{Observations}

\subsection{Near-Infrared Camera: CISCO}

A near-infrared imaging and spectroscopic instrument called CISCO 
(Cooled Infrared Spectrograph/Camera for OHS) was used from early 
April to mid-June 1999 at the Cassegrain focus of the Subaru 
Telescope.  The detailed description of CISCO has been reported by 
Motohara et al.  (1998).  Major specifications are listed in 
table 1. 


\begin{table*}[h]
\begin{center}
Table~1.\hspace{4pt}Specifications and performance of CISCO.\\
\end{center}
\vspace{6pt}
\begin{center}
\begin{tabular}{cc}
\hline\hline
Item & Description \\
\hline
F.O.V. & $2^{\prime}$ \ $\times$ \ $2^{\prime}$ \\
Pixel scale  & 0$.\hspace{-2pt}''$116 pix$^{-1}$ \\
Filter & $z$, $J$, $H$, $K^{\prime}$, $K$, NBF2.04, NBF2.12, NBF2.25 \\
Grism & $zJ$, $JH$, $K$ \\
Limiting magnitude & 22.6 mag ($K^{\prime}$-band, 1 hour, S/N=5) \\
\hline
\end{tabular}
\end{center}
\end{table*}

The filters used in the present deep survey are $J$ (1.16--1.32 
$\mu$m) and $K^{\prime}$ (1.96--2.30 $\mu$m) bands.  A fixed exposure 
time was employed throughout observations, namely, 40 sec in the $J$ 
band and 20 sec in the $K^{\prime}$.  The exposure times are shorter 
than the saturation level (about 30\%) of the detector readout system, 
but we adopt them throughout the present survey to secure uniformity 
in the dithered multiple exposure strategy we adopted (see the 
following subsection).

\subsection{Field Selection}

Several years ago we chose two regions suitable for a very deep survey, 
one each in the northern and southern hemispheres, and we designated 
them as the Subaru Deep Field (SDF).  The selected area in the present 
near-infrared survey is one of the pre-determined blank sky regions 
near the north Galactic pole (NGP), as shown in figure 1. 
The center coordinates of the observed area of SDF-NGP are, 
RA(2000)=13$^{\rm h}$24$^{\rm m}$21$.\hspace{-2pt}^{\rm s}$38 and 
DEC(2000)=+27$^{\circ}$29$^{\prime}$23$.\hspace{-2pt}''$0. 
In selecting the SDF survey region, we placed a requirement that a 
reasonably bright star be located close to the area to serve as a 
reference star for possible adaptive optics-based observations in the 
near future.  The principles and criteria for selecting the SDF 
regions are described below:

i) We chose an independent deep survey region because the nature 
(appearance) of the Universe may have different characteristics from 
one direction to another.
ii) The spatial resolution achieved by the Subaru-CISCO combination 
has proved to be nearly 0$.\hspace{-2pt}''$3 or even better under the 
best conditions, and about 0$.\hspace{-2pt}''$45 on average, which 
potentially offers an excellent opportunity to probe remote faint 
galaxies in the high-redshift Universe.  We therefore hope to obtain 
the deepest survey data in the near-infrared, especially in the 2 
$\mu$m region by taking long enough observations.
iii) The HDF is at higher airmass at Mauna Kea than that of the SDF. 
This factor is significant in the latter half of nights from April to 
June, when the present survey was performed.  Note that, although the 
limiting magnitude of HST-NICMOS is extremely high compared with any 
ground-based near-infrared imaging, the sensitivity in the $K$ or 
$K^{\prime}$ band is higher in the ground-based ones due to larger 
telescope aperture.  The sensitivity of optical bands by Subaru Telescope 
in terms of {\it total magnitude} is expected to be nearly comparable 
or even higher ($R$-band, for example), despite lower spatial resolution.
iv) We should have a reference star nearby for AO-based observations.  
With the AO system, the spatial resolution is much higher than that of 
HST-NICMOS in the $H$ and $K$ bands, and the sensitivity is comparable 
if the exposure time is the same.
v) In selecting SDF, we posed additional requirements, namely, low 
Galactic HI column density, no nearby bright stars and galaxy (except 
for the reference star for AO), and no known nearby cluster of galaxies. 

\subsection{Method of Dithering Observations}

We employed the 8-position dithering method in both bands.  In the 
$K^{\prime}$-band case, 12 consecutive frames were taken at each 
position before offsetting the telescope towards the next position.  
The order of 8 positions was chosen so as to make a diamond pattern 
with a diameter of 8$''$-12$''$.  Thus we took 96 frames of 20 sec 
exposures in a single set of dithering observations, resulting in a 
total on-source integration time of 1920 sec.  The entire process of 
taking 96 exposures is organized by an `abstract command' (kind of a 
macro command) issued by the observer workstation which dispatches 
individual commands to the telescope control system, such as the 
auto-guiding system, the infrared camera controller, and the telescope 
itself.  Normally, we made observations in one of the two photometric 
bands, $J$ or $K^{\prime}$, in one night throughout the SDF project.

When the seeing conditions were fairly good, we preferentially 
implemented $J$-band observations in the fashion described above.  The 
same integration time of 1920 sec in a set of dithering observations 
was secured, but by adopting single exposure time of 40 sec with 6 
consecutive exposures at one position instead of 12 in the $K^{\prime}$ 
band case, in view of the lower sky background level of the $J$ band.   
The positional accuracy of each offset movement relative to the origin 
of coordinates, when the direction is put back, has proved to be on 
the order of 0$.\hspace{-2pt}''$05 under normal conditions, but 
sometime it is as large as 0$.\hspace{-2pt}''$2, as measured by the 
center-of-gravity fluctuation of a bright object on the recorded 
frames.  We determined offset values of each frame by the measured 
coordinates of a particular star in the frames, not by referencing the 
recorded FITS data.

\subsection{Log of Observations} 

Observations of the SDF survey started on April 3 and 
ended on June 22, 1999, using available nights for CISCO observations 
during the telescope commissioning phase. Since the SDF survey was 
supported by the Subaru Project Office as a priority project during this 
period, we devoted almost all nights with observing conditions better 
than average to the SDF work. The observing log is presented in 
table 2. 
Note that the nominal date in the table corresponds to the starting 
UT time of each night. 


\begin{table*}[t]
\begin{center}
Table~2.\hspace{4pt}Observing log.\\
\end{center}
\vspace{6pt}
\begin{center}
\begin{tabular}{ccccccc}
\hline\hline
Date  & Band & Exposure   & Seeing & No. of frames & Total exp. & Comment \\
(UT)  &   \  & time (sec) & FWHM($^{\prime\prime}$) & \  & time (sec) & \ \\
\hline
 [2/27/99] & $K'$ &    20   &   0.35-0.80  &     96 x 3     &      5760  &    test observation \\
   4/ 3/99 & $K'$ &    20   &   0.40-0.70  &     96 x 2     &      3840  & \\
   4/ 4/99 & $K'$ &    20   &   0.35-0.60  &     96 x 2.75  &      5280  & \\
   4/ 9/99 & $K'$ &    20   &   0.70-1.10  &     96 x 0.5   &       960  & \\
   4/25/99 & $K'$ &    20   &   0.60-1.20  &     96 x 4     &      7680  &    flanking field 1-4 \\
   4/25/99 & $K'$ &    20   &   0.80-1.20  &     96 x 2     &      3840  &    flanking field 5-6 \\
   4/29/99 & $J$  &    40   &   0.45-0.65  &     48 x 2.5   &      4800  & \\
   4/30/99 & $J$  &    40   &   0.35-0.50  &     48 x 2.75  &      5280  & \\
   5/ 2/99 & $J$  &    40   &   0.40-0.65  &     48 x 6.5   &     12480  & \\
   5/ 6/99 & $J$  &    40   &   0.45-0.65  &     48 x 2     &      3840  & \\
   5/ 7/99 & $J$  &    40   &   0.35-0.50  &     48 x 2     &      3840  & \\
   5/ 8/99 & $K'$ &    20   &   0.35-0.55  &     96 x 1     &      1920  & \\
   5/11/99 & $K'$ &    20   &   0.25-0.60  &     96 x 6.5   &     12480  & \\
   5/27/99 & $K'$ &    20   &   0.20-0.30  &     96 x 3.5   &      6720  & \\
   6/ 6/99 & $J$  &    40   &   0.35-0.80  &     48 x 3     &      5760  & \\
   6/ 7/99 & $K'$ &    20   &   0.35-0.55  &     96 x 2     &      3840  & \\
   6/ 7/99 & $J$  &    40   &   0.45-0.75  &     48 x 2     &      3840  & \\
   6/ 9/99 & $J$  &    40   &   0.35-0.40  &     48 x 2     &      3840  & \\
   6/20/99 & $K'$ &    20   &   0.50-1.20  &     96 x 2     &      3840  &   flanking field 7-8 \\
   6/22/99 & $K'$ &    20   &   0.40-0.65  &     96 x 1     &      1920  &   flanking field 8 \\
\hline 
\end{tabular}
\end{center}
\end{table*}

During this period, the best seeing data were recorded on May 26; a 
couple of frames of a field star with 20 sec exposures had PSFs of 
$\sim 0.20^{\prime\prime}$ at FWHM. Although the individual frames 
have different image resolutions, mostly attributed to the day-to-day 
variations in seeing conditions, we just accumulated all the acquired 
frames without considering the difference in seeing in the present 
analyses.  Thus the present near-infrared SDF survey has ended up with 
the target dedicated time of 12.1 hours in the $J$ band and 9.7 hours 
in the $K^{\prime}$ band.

\section{Data Reduction and Source Detection}

Using all the frames taken in a set of automatic dithering 
observations, we first created a sky frame out of 96 dithered 
exposures by the standard median sky method.  In this procedure, each 
frame is first divided by the flat frame and is also corrected by a 
bias frame.  The flat frame is produced from a large number of sky 
frames by a method similar to the co-adding method used for creating 
the standard median sky but using all the frames obtained in the 
long-term SDF observations.  In generating the sky frame, the 
contribution of individual stars was maximally reduced by masking all 
the discernible stars.  Then, we co-added all the sky-subtracted 
frames in which the offsets of the shift-and-add procedure had been 
determined at a sub-pixel level from the bright sources in each frame.  
The pixel values with scatters exceeding 3 $\sigma$ were rejected, and 
therefore possible cosmic ray events or rare pixel events due to 
unstable behavior were excluded.

The reduced $J$ and $K^{\prime}$ band images are shown in figures 2 
and 3, respectively, and the color composite created from these two 
bands is presented in figure 4.  Throughout these images, the center 
position is 13$^{\rm h}$24$^{\rm m}$21$.\hspace{-2pt}^{\rm s}$38 in RA, 
and +27$^{\circ}$29$^{\prime}$28$.\hspace{-2pt}''$3 in Dec at the 
epoch of 2000.0 as determined by the astrometric measurement based on 
the position of a bright reference star located in the northeastern 
corner of the SDF field, which is just trimmed out in these figures.  
The unvignetted area presented in figure 2 is 118$'' \times$ 
114$''$ with a pixel scale of 0$.\hspace{-2pt}''$116 per pixel.  The 
stellar image sizes of figures 2 and 3 are 0$.\hspace{-2pt}''$45 in FWHM  
for the $J$ band, and 0$.\hspace{-2pt}''$35 for the $K^{\prime}$ band. In 
producing the combined color image of figure 4, we applied a smoothing 
filter to the $K^{\prime}$ image to match the image size in $J$ of 
0$.\hspace{-2pt}''$45 at FWHM.

For source detection and photometry on the reduced SDF frames, we 
employed the routine called SExtractor developed by Bertin and Arnouts 
(1996).  Before applying it, images of both bands were smoothed out by 
a Gaussian filter that makes image resolution 0$.\hspace{-2pt}''$55 at 
FWHM because we have learned that filtered images give optimal source 
detection capability with less spurious source detection by iterative 
trials of SExtractor.  
We define the detection threshold as the 1.50 $\sigma$ level of 
surface brightness fluctuation of the sky, which corresponds to 25.59 
mag arcsec$^{-2}$ in the $J$ band and to 24.10 mag arcsec$^{-2}$ in 
the $K^{\prime}$ band.  If an assemblage of at least 18 pixels, which 
are connected to each other, has an excess signal over the thresholds, 
we regarded it as positive detection.  We have thus defined the 
isophotal magnitude by integrating the signal within the region that 
exceeds the threshold level.  In addition, we have listed total 
magnitude as well as aperture magnitude in our primary catalog.  

The total numbers of sources detected and cataloged through the above 
procedures are 911 and 939 in the $J$ and $K^{\prime}$ bands, 
respectively.  Note that the catalog may contain spurious sources due 
to the effect of noise.  Regardless of the reliability of each source, 
we have accomplished photometric calibration of the data by a 
reference star in the same frame whose brightness has been determined 
by measuring standard stars: FS23 and FS27, selected from the UKIRT 
faint star catalog (Casali \& Hawarden 1992).  The resultant calibrated 
catalog will be reported elsewhere when optical data are obtained.  

To obtain a diagram of number count versus total magnitude, we have to 
assess the spurious detection of sources due to the effect of 
statistical noise and also the error of source photometry.  For noise 
evaluation, we first created a reference frame (artificial blank sky) 
using SDF raw frames, but with all traces of detected objects removed.  
To ensure removal of much fainter objects from the reference frame, we 
also referred to the co-added object frames for identification.  The 
final blank sky frames in both bands were thus obtained by co-adding 
all the frames without adjusting the dithering offset.  Then, by 
applying SExtractor to the sky frame, we evaluated the rate of 
spurious source detection, and thus obtained the correction factor for 
the raw result.  It was found that spurious source detection tends to 
affect the source count discussed below, but only in the S/N $<5$ 
range.

In the next step, a number of artificial objects (mock galaxies) with 
a wide range of brightness as well as source size were embedded in the 
real image frames.  The brightness distribution of the objects in the 
faint region are assumed to be represented by a slope of $\sim 0.23$ 
(derived from the raw data with S/N $>5$).
In the course of this calculation, the SExtractor software was applied 
multiple times, to establish a relationship between the input 
photometric brightnesses and the measured ones.  The results of 
simulations are shown in figure 5, where each dot represents the 
measured magnitude for an input object of a given magnitude.  The 
relation is expressed by a matrix-type operator.  This method has been 
employed in galaxy count studies (e.g., Smail et al.  1995; Minezaki 
et al.  1997).  Based on the matrix, we can evaluate errors in 
photometry, and also correct the raw number count data using the 
completeness curves as shown in the lower panel of figure 5.

The source counts against magnitude, with and without the correction 
for completeness, are plotted in figures 6 and 7, where data points of 
total, isophotal, and aperture magnitudes are presented for 
comparison.  As seen from these figures, the correction becomes 
significant at magnitudes larger than 25.5 mag in $J$ and 24 mag in 
$K^{\prime}$.  If we define a S/N of 5 for definite detection, the 
limiting magnitudes are 25.1 mag in $J$ and 23.5 mag in $K^{\prime}$, 
with the number of sources being 385 and 350, respectively.  The 
magnitudes corresponding to detection completeness of 50\% are 24.4 
mag in $K^{\prime}$ and 25.8 mag in $J$.

\section{Results and Discussion}

\subsection{Corrected Number Counts}

Corrected number counts in both the $J$ and $K^{\prime}$ bands are 
tabulated in table 3.  In figures 8 and 9, we plot them for $J$ band 
sources and for $K^{\prime}$ band sources, respectively.  
Also plotted are those 
of other surveys taken from the literature.  In the table and figures, 
our data refer to {\it total magnitude}, while some of the other data 
points are defined by {\it aperture magnitude}.  Source sizes in 
the faintest magnitude 
range are very small, and in general, smaller than the aperture 
adopted in most of the aperture photometry.  This means that the 
aperture magnitude is virtually the same as total magnitude, since the 
aperture is normally taken to be larger than the seeing size.
In this first report of the SDF survey program, we shall concentrate 
mostly on the sources having relatively high S/N-ratios, S/N of 
$\sim$5 or more.  We will prepare a separate paper in which galaxy 
counts, colors, and morphologies at the faint end will be examined.  

\begin{table*}[t]
\begin{center}
Table~3.\hspace{4pt}{Corrected number counts}\\
\end{center}
\vspace{6pt}
\begin{tabular*}{\textwidth}{@{\hspace{\tabcolsep}
\extracolsep{\fill}}p{6pc}ccccc}
\hline\hline\\ [-6pt]
Magnitude& \multicolumn{2}{c}{$J$-band}& \multicolumn{2}{c}{$K'$-band}\\
[4pt]\cline{2-5}\\[-6pt]
               & count & error & count & error\\[4pt]\hline\\[-6pt]
16.25  & 0.000e+00 & 0.000e+00 & 1.918e+03 & 1.931e+03\\
16.75  & 0.000e+00 & 0.000e+00 & 1.358e+02 & 5.118e+02\\
17.25  & 0.000e+00 & 0.000e+00 & 5.635e+03 & 3.293e+03\\
17.75  & 3.231e+01 & 2.500e+02 & 5.755e+03 & 3.338e+03\\
18.25  & 1.949e+03 & 1.942e+03 & 7.721e+03 & 3.869e+03\\
18.75  & 3.851e+03 & 2.725e+03 & 9.653e+03 & 4.326e+03\\
19.25  & 5.769e+03 & 3.354e+03 & 1.346e+04 & 5.100e+03\\
19.75  & 3.996e+03 & 2.776e+03 & 1.915e+04 & 6.111e+03\\
20.25  & 1.314e+04 & 5.044e+03 & 2.885e+04 & 7.513e+03\\
20.75  & 1.368e+04 & 5.162e+03 & 4.757e+04 & 9.647e+03\\
21.25  & 2.296e+04 & 6.705e+03 & 5.145e+04 & 1.006e+04\\
21.75  & 3.099e+04 & 7.789e+03 & 7.073e+04 & 1.187e+04\\
22.25  & 4.357e+04 & 9.255e+03 & 8.419e+04 & 1.307e+04\\
22.75  & 5.835e+04 & 1.070e+04 & 1.066e+05 & 1.488e+04\\
23.25  & 7.499e+04 & 1.224e+04 & 1.487e+05 & 1.873e+04\\
23.75  & 8.759e+04 & 1.333e+04 & 1.962e+05 & 3.129e+04\\
24.25  & 1.131e+05 & 1.524e+04 & 2.548e+05 & 5.648e+04\\
24.75  & 1.615e+05 & 1.905e+04 & 3.421e+05 & 1.031e+05\\
25.25  & 2.160e+05 & 2.949e+04 & 4.793e+05 & 2.214e+05\\
25.75  & 2.624e+05 & 5.623e+04 &    ---    &    ---   \\
26.25  & 3.451e+05 & 1.227e+05 &    ---    &    ---   \\
[4pt]\hline
\end{tabular*}
\end{table*}

Here it should be noted that the contribution of stars to the number 
counts is estimated by a simulation, and that the point sources (14 
objects in the SDF field have been identified) have been excluded.  
The actual procedure of the simulation by which a criterion to 
identify stellar objects will be described in a later section.
Some quasi-stellar objects (QSOs) may also be included in the thus 
identified stellar objects.  However, their contribution to the number 
count should be smaller than that of stars in view of the result found 
by Huchra and Burg (1992), who reported that the fraction of all the 
types of Seyfert galaxies at the absolute magnitude limit of $-20.0$ 
is about 1.3\% in the CfA redshift survey.  Brighter objects with a 
stellar appearance are expected to be less abundant than Seyfert 
galaxies.

In figure 10, we plot the galaxy number counts against the AB 
magnitude scale to compare the slope as well as absolute number 
density of different photometric bands obtained so far.  Here the 
HST-NICMOS $J$- and $H$-band data are taken from Yan et al.  (1998) 
and Thompson et al.  (1999).  In these figures, published data of 
galaxy number counts by the authors are also plotted.
The spatial number density of galaxies in the SDF appears to be a good 
match with that given by Yan et al.  (1998), who presented the 
$H$-band number count obtained with HST-NICMOS operating in the 
so-called parallel mode.  It is also consistent with Thompson 
et al. (1999) at the faint end (at $H=23.5-26$ AB mag) but 
deviates much at magnitudes brighter than 22.5 mag possibly due to 
lower statstics. 
In contrast, the absolute number densities given by some of the past 
ground-based surveys differ substantially from that of the present SDF 
survey in the magnitude range $\geq$ 23 AB mag.  It would still be 
necessary to increase survey areas as well as to obtain higher S/N 
ratios to examine the possible structural inhomogeneity of the 
Universe.  Nevertheless, the SDF area is so far the largest among deep 
near-infrared surveys, and therefore should be a more accurate
representation of the global distribution of galaxies in the Universe.
Another important result derived from figures 8, 9, and 10 is that the 
near-infrared color $J-K^{\prime}$ is almost constant at least in the 
AB mag range from 22 to 25.5 mag, and the median color is $\sim 1.4$, 
as described next.

\subsection{Extragalactic Background Light (EBL)}

As for the slope of the galaxy count in the $K'$ band, we have to be 
careful in interpreting it because we have assumed a slope of 0.23 in 
the model source number count in the simulation process.  This affects 
the result at the faint end, although the slope is iteratively 
corrected and has converged at 0.23.  It should also be noted that 
this slope is derived by applying the correction factor of completeness, 
in which we have assumed, for simplicity, that these faint galaxies 
are point sources, since aparent sizes of detected faint sources are 
small enough. 
  Figure 11 shows the contribution to the EBL in the $K'$ band as a 
function of apparent magnitude.  This figure shows that we have 
already resolved the EBL consisting of discrete galaxies in this band.  
The optical galaxy counts in the HDF have also shown such a 
signature (see Pozzetti et al.  1998).  The present result of the SDF 
in the near-infrared band gives clear evidence that the bulk of EBL is 
contributed by an integration of fairly bright discrete sources, 
consistent with a similar diagram shown by Pozzetti et al. (1998), 
who compiled available $K$-band data. \\

Now we can evaluate surface brightness as EBL by integrating 
individual sources in a wide magnitude range of $10 \leq K' \leq 25$. 
The estimate using our data as well as published count data is 
$\sim 5.1 \times 10^{-20}$ erg cm$^{-2}$s$^{-1}$Hz$^{-1}$sr$^{-1}$ 
in the $K'$ band.  
The major contribution to EBL is made by fairly bright galaxies of 
about $K' \sim$ 15-20 mag.  Note that the contribution of sources 
fainter than $K' = 25$ mag to surface brightness is small, since the 
slope is no more than 0.23.  Extrapolation both into the faint and 
bright end adds at most $\sim$ 5\% of the above value of the EBL flux. 
It should be noted that these estimates are based on the completeness 
correction that assumes that all galaxies are small in size.  Since 
the completeness correction may be larger than this, if there are 
galaxies with larger spatial extents, and thus with lower surface 
brightnesses. Then the above estimate may be an underestimate of 
the $K'$-band EBL. However, as seen in figure 11, the bulk of EBL 
comes predominantly from relatively bright galaxies 
at $K' \sim$ 15-20, so that we consider that the uncertainty in the 
counts at the faintest magnitudes in the SDF does not significantly 
change the above estimate of the EBL.\\

\subsection{Near-infrared Color and Morphology of Faint Galaxies} 

Figure 12 shows the $J-K^{\prime}$ color vs. $K^{\prime}$ magnitude 
diagram of near-infrared sources.  Here we omitted sources that have 
been assigned as stellar objects.  Since the photometric aperture of 
total magnitude in the $J$ band is not always the same as for the $K'$ 
band, it is necessary to define a color with the same aperture of the 
two bands.  Therefore we have calculated colors of smaller sources 
in figure 12 in terms of photometric magnitudes defined by 10 pixels, 
i.e., a 1$.\hspace{-2pt}''$16 diameter.  
Even if an object is picked up only in the $J$ band and is not detected 
by our criteria in the $K'$ band, a $K'$ photometric brightness 
is artificially given by measuring encircled intensity with the same 
aperture as that defined in the $J$ band, thus providing the 
$J-K^{\prime}$ color of the sources. 

The median color is shown by the filled squares in figure 12, while 
the thin lines represent color magnitude curves for four categories of 
galaxy, E/S0, Sbc, Scd, and Irr, all of which being drawn by applying 
only K-correction (Coleman et al.  1980; Yoshii \& Takahara 1988).  
Pozzetti et al. (1996) presented a study of the pure luminosity 
evolution (PLE) model to examine faint galaxy count data from the U- 
to the K-band in which luminosity evolution as well as mild spectral 
evolution are incoporated, and they have shown that simple PLE 
models are in general considered as baseline models of faint galaxy 
counts. 
However, as they also noted, the discrepancy between PLE models and 
the $K$ band data is more significant than optical bands. 
One may notice that, as shown in figure 12, the median color of 
sources is fairly constant up to about $K'$=22 mag, and then gets 
slightly redder as the brightness gets fainter, although the standard 
deviation is large.  As for the color of faint galaxies, Saracco et al.  
(1999) argued that the median $J-Ks$ color of galaxies gets redder 
from 1.1 to 1.5 up to $Ks$=19 mag, and then it tends to be somewhat 
bluer in the fainter magnitude range. Such a trend is not necessarily 
inconsistent with figure 12, but it is note worthy that the color is, 
by and large, constant at least in the magnitude range where 
the selection effect is still small (SN$>$5 for both bands), but with 
a significant statistics.  In order to interperete observed 
galaxy counts as well as the color 
magniture relation, it is necessary to introduce galaxy evolution 
models in which different formation epochs ($z_{f}$) for different 
galaxy categories are presumed. Such quantitative analyses for the 
present $J-K'$ color vs. luminosity relation will be discussed 
in our forthcoming paper. \\

In figure 12, sources with $J-K^{\prime}$ colors redder than 2.5 are 
discerned, where 4 objects out of 9 are not detected in the $J$ band.  
These large $J-K^{\prime}$ color objects in the faint region at about 
$K'$ = 23 mag are likely to be remote galaxies as judged from the 
apparent spatial extent as well as the derived ``stellarity'' index 
determined by SExtractor.  Some of the reddest objects in the SDF 
survey images are tabulated in table 3 and aldo shown in figure 13.  
The listed objects are relatively bright in $K'$ ($<$22.5 mag) and 
have $J-K^{\prime}$ colors equal to or larger than 2.8.  The object 
on the left appears to be a merger system, and the third one from the 
left may be representing an interacting system, 
although it is possible that they could just appear 
as close neighbors.

\begin{table*}[h]
\begin{center}
Table~4.\hspace{4pt}$K'$-magnitude and color of extremely red objects.\\
\end{center}
\vspace{3pt}
\begin{center}
\begin{tabular}{cccc}
\hline\hline
ID Number$^{\ast}$ & Position$^{\dagger}$ (2000.0) & $K'$ & $J-K'$  \\
\hline
1 &13$^{\rm h}$24$^{\rm m}$22$.\hspace{-2pt}^{\rm s}$38 +27$^{\circ}$29$^{\prime}$49$.\hspace{-2pt}''$5 & 20.91 ($\pm$0.05) & 2.97 ($\pm$0.14) \\
2 &13$^{\rm h}$24$^{\rm m}$22$.\hspace{-2pt}^{\rm s}$39 +27$^{\circ}$29$^{\prime}$01$.\hspace{-2pt}''$9 & 22.03 ($\pm$0.09) & 3.65 ($\pm$0.40) \\
3 &13$^{\rm h}$24$^{\rm m}$21$.\hspace{-2pt}^{\rm s}$16 +27$^{\circ}$29$^{\prime}$01$.\hspace{-2pt}''$9 & 21.99 ($\pm$0.05) & 2.81 ($\pm$0.20) \\
4 &13$^{\rm h}$24$^{\rm m}$22$.\hspace{-2pt}^{\rm s}$84 +27$^{\circ}$30$^{\prime}$08$.\hspace{-2pt}''$4 & 22.31 ($\pm$0.14) & 4.12 ($\pm$1.04) \\
\hline
\end{tabular}
\end{center}
$^{\ast}$ID numbers assigned to objects are from left to right in figure 13. \\
$^{\dagger}$Astrometry of these objects was made by the coordinates of an HST guide star \\
found in the flanking field. The estimated accuracy is $\pm$0$.\hspace{-2pt}''$15.\\
\end{table*}

Extremely red objects with unprecedentedly large $J-K'$ colors located 
in the faint magnitude domain are currently given special attention in 
connection with galaxy formation in the earliest epoch.  Dickinson et 
al. (2000) found an unusual infrared object in the HDF North field 
detected only in the $H$ and $Ks$ bands with no detectable fluxes 
shorter than the J band.  It has the $H-K$ color of about 1.2 with the 
$J-H$ color limit of nearly 3.  They discuss three possible 
interpretations, which are, i) a dusty z $>2$ galaxy, ii) an old 
elliptical at z $>3$, or iii) a z $>$10 Lyman break galaxy.  Similar 
objects were reported by Yahata et al.  (2000), who list possible 
extremely high redshift galaxies in the HDF South NICMOS field.  Nine 
objects appear to have a break between 1 to 2 $\mu$m.  Redshifts of 
the source have been derived from $U$- to $K$-band data by a method of 
photometric redshift determination that spans from z = 7.66 to 15.45.

Since we have so far not acquired any data in the optical bands for 
the extreme SDF objects, we cannot infer photometric redshifts on a 
firm basis.  Nevertheless, infrared color as large as $J-K'\geq$ 2.8 
suggests these objects belong to the same population of the HDF-South
NICMOS objects having redshift z $\geq$ 10.  It is however necessary 
to get optical band photometric information, as well as spectroscopic 
data with highly sensitive instruments such as the OHS, for further 
examination of these objects.  We can determine whether they are 
moderately redshifted (z$\sim2-4$) ellipticals or unprecedentedly 
high-redshift galaxies. 

\subsection{Faint Stellar Populations in the Galaxy}

In order to extract stars from the table of SDF sources detected at 
S/N $> 5$, we have developed a procedure of star-galaxy separation 
based on a detectivity test using mock stars.  It is similar to the 
completeness test of the source count in conjunction with the 
SExtractor.  The criteria for identifying stars from SDF sources are 
basically expressed by the following two conditions.  The first one is 
that the FWHMs are smaller than 0$.\hspace{-2pt}''$47, and 
0$.\hspace{-2pt}''$34 for bright sources, in the $J$ and $K'$ bands, 
respectively.  In addition, based on the simulation, it is found that 
the limiting sizes should be corrected in the fainter source region, 
by adding terms: i.e., $10^{(m-m_{e})/2.5}$, where m is the magnitude 
of the source, and $m_{e}$= 26 and 25 mag for the $J$ and $K'$ bands, 
respectively.  The second condition is that the stellarity index which 
is given by SExtractor is larger than 0.8.  By adopting these 
criteria, 14 objects are classified as stars on a fairly firm basis.  
Naturally, possible stellar objects with stellarity indexes larger than 
0.8 could be excluded due to the first criterion.  See Nakajima et al.  
(2000) for the detailed procedure of galaxy-star separation we have 
developed.  The completeness of identification of stars has also been 
estimated with this simulation, which is about 60\% for sources 
brighter than 24 mag in the $J$ band, or 23 mag in the $K'$ band.  We 
plot the SDF objects in the FWHM vs.  $J-K'$ color diagram as shown in 
figure 14, where objects satisfying the above criterion are marked by 
open star symbols.

It is interesting to note that several relatively bluer ($J-K'$ $\sim$ 
0.6) stellar objects appear to concentrate in the lower left: a 
well-confined region on the diagram.  These are presumably {\it 
extreme M subdwarf} (ESD) stars as classified by Leggett, Allard \& 
Hauschildt (1998).  The extreme M subdwarfs are supposed to be members 
of the Galactic halo and have very low metallicity.  They are listed 
in the tables of low mass stars presented by Leggett, Allard, \& 
Hauschildt (1998), and are characterized by effective temperatures of 
about 3000 K and by masses of 0.09 $\sim$ 0.15 $M_\odot$.  From their 
photometric data it is seen that the brightness range of ESD stars is 
from $K$=11 to 15 mag and that the $J-K'$ color is 0.65 $\pm$ 0.15.  
In view of these, some 10 stars in the confined region of figure 14 
are most probably extreme M subdwarfs.  Since their absolute 
magnitudes span from $M_{J}$=9.5 to 10, we should have reached about 8 
kpc to capture ESD stars by the present observations with $K'$ 
band-limiting magnitudes of $\sim$ 24.5 mag.  The estimated volume 
density due to these stars is 2.5 $\times$ 10$^{-4} M_\odot$ 
pc$^{-3}$.  Even if the density of stars is extended uniformly to 30 
kpc, the total mass is a little lower than 10$^{11}M_\odot$, so that 
it could not be the majority of Galactic dark matter.  

Another group of stellar components, significantly redder than M 
subdwarfs, is noticed in figure 14 in the $J-K'$ color range from 1 to 
1.5, .  It is likely that they correspond to the L-type dwarfs defined 
by Kirkpatrick et al.  (1999), who have found very red stellar objects 
with $J-Ks$ colors from 1.3 to 2.1 (practically the same color as the 
$J-K'$ color).  They have claimed that at least one third of L-type 
stars shows lithium absorption, and that they are definitely in the 
category of brown dwarfs.

On the other hand, note that objects with $J-K'$ color of $\sim$0 and 
small FWHM values are plotted in figure 14.  In fact, they had a fairly 
large stellarity index ($>$0.8) in the $J$ band, but were dropped from 
identification due to faintness in the $K'$ band.  The color is 
consistent with a T-type brown dwarf, GL229B (Nakajima et al.  1995; 
Kirkpatrick et al.  1999).  However, it is necessary to 
prove show the nature of these L-type and T-type candidates through
future multi-wavelength photometric and spectroscopic observations.
Related discussions on the stellar members in SDF will be presented in 
Nakajima et al.  (2000).

Finally, it is worth noting that the SDF sources have a connection to 
old, cool white dwarfs, which have recently drawn attention because 
they might account for most of the hidden baryonic mass of the Galaxy.  
Hodgkin et al.  (2000) reported a cool white dwarf showing an 
extraordinary spectrum affected by the collision-induced absorption by 
hydrogen molecules, i.e., very red in the optical region but extremely 
blue at wavelengths longer than 1 $\mu$m, with a $J-K \sim -1.4$.  
In addition, Harris et al.  (2000) identified LHS3250 as a very cool 
white dwarf with a $J-K$ of $-0.86$.  A couple of old white dwarf 
candidates were also found with HDF frames taken at a 2-year interval, 
which are believed to be halo members as inferred from proper motion 
data (Ibata et al.  1999). These candidates were spectroscopically 
shown to be white dwarfs (Ibata et al.  2000).

These findings support the idea that very old white dwarfs could be 
responsible for a substantial portion, but not all, of the Galactic 
dark matter.  In this connection, we have few stellar objects (only 
one below $J$=25.0) having stellarity index of 0.8 or larger, and with 
very blue color.  This scarcity is consistent with the rough 
estimate by Hodgkin et al.  (2000), who gave $\sim$7 $\times$ 
10$^{-3}$ pc$^{-3}$ as a number density of white dwarfs with a typical 
mass of 0.5 $M_{\odot}$ in the solar neighborhood.  If the absolute 
$J$ magnitude, M$_{J}$ of white dwarfs is assumed to be the same 
as WD 0346+246 of Hodgkin et al.  (2000) with the detection limit 
being m$_{J}$ = 24.75, 
then the viewable radial distance would be about 750 pc.  The survey 
volume of the 2$^{\prime}$ $\times$ 2$^{\prime}$ SDF field becomes 
$\sim$ 50 pc$^{3}$, which is about a half of the volume of 109 pc$^{3}$ 
surveyed by Hodgkin et al.

In order to put a definite constraint on the number density of the white 
dwarf population, it is necessary to increase the survey region and 
also to perform an optical survey.  Since the magnitudes in the optical 
bands are roughly the same as in the $J$ band ($\sim$25 mag), such dwarf 
stars will be seen as optical images, by which the peculiar photometric 
color could be determined.

\section{Summary}

A deep near-infrared survey with the newly commissioned Subaru 
Telescope has been reported.  We present two color data of $J$ and 
$K'$ bands as deep images, as well as diagrams, of galaxy number 
count vs.  magnitude, of color vs.  magnitude, and of size vs.  color.  
i) It is found that slopes of galaxy number count plotted 
against the AB mag scale in both the $J$ and $K'$ bands are about 0.23 
in the 22 to 26 AB mag range, which remain the same up to the faint 
end without a significant change. 
ii) From this result, we argue that the integrated surface brightness 
of faint SDF sources does not make an appreciable contribution to the 
extragalactic background light (EBL), or in other word, that the total 
EBL contributed by galaxies up to the faint end has nearly comletely 
been resolved. However, measurements of diffuse EBL at this wavelength, 
if performed with enough precision, may pose a crucial cosmological 
issue regarding the light sources other than individual galaxies. 
iii) The color-magnitude diagram shows a fairly constant $J-K'$ 
color ($\sim$ 1.5) up to about $K' \sim$ 23 mag, although a scattered 
distribution, especially toward redder color, is noticed. 
vi) Intriguing objects with extremely red $J$-$K'$ colors are found in 
the SDF region. The listed objects are reltively bright (SN$>$5), and 
thus are safely classified as galaxies. The possibility of very high 
redshift objects, that is, candidates of Lyman break galaxies, are 
briefly discussed.   
v) A certain number of stellar sources have been identified, most 
of which are supposed to be M subdwarfs having colors of $J-K'\sim$ 0.6.  
These are considered to be located at a median distance of about 2 kpc 
and are expected to provide samples for further studies of stars of 
this class for the purpose of examining a luminosity function 
as well as a contribution to the Galaxy mass.  Brown dwarfs of the 
T-type may also be included in the detected source list.  This should 
be confirmed by spectroscopic measurements.\\

  Finally, it should be noted that the present near-infrared SDF survey 
work is the first step of the Subaru Deep Survey project planned to 
perform using other facility instruments of Subaru Telescope.  For 
instance, follow-up observations of optical deep imaging with the 
optical spectrograph/camera called FOCAS (Kashikawa et al. 2000) will 
provide essential data to determine photometric redshifts of detected 
near-infrared sources. Objects as faint as $H$=21.5, or $J$=22 mag 
are well within the feasible range of spectroscopic observations 
with the OH-airglow suppression spectrograph (OHS), which has a 
unique capability of obtaining 1.1 to 1.8 $\mu$m spectra of 16 
objects within a 3$'$ field simultaneously, giving crucial information 
about the SDF sources. \\

\vspace{1pc}\par
We deeply appreciate the devoted support of the Subaru Telescope
staff for this project.  We also thank the engineering 
staffs of Mitsubishi Electric Co. and  Fujitsu Co. for 
technical assistance during the observations.  The authors would like 
to acknowledge helpful discussion about stellar components with 
Tadashi Nakajima and Takashi Tsuji. 


\section*{References} 

\re Ben\'{\i}tez N., Broadhurst T., Bouwens R., Silk J., Rosati, P.
  1999,  ApJ, 515, L65
\re Bershady M.A., Lowenthal J.D., Koo D.C. 1998, ApJ, 505, 50
\re Bertin E., Arnouts S. 1996, AA, 117, 393
\re Casali M., Hawarden,T. 1992, JCMT-UKIRT Newsletter No.4, 33
\re Coleman G.D., Wu C.-C., Weedman D.W. 1980, ApJS, 43, 393
\re Dickinson M., Hanley C., Elston R., Eisenhardt P.R., 
  Stanford S.A., Adelberger K.L., Shapley A., Steidel C.C., 
  Papovich C., Szalay S., Bershady M.A., Conselice C.J., 
  Ferguson H.C., Fruchter A.S. 2000, ApJ, 531, 624
\re Gardner J.P. 1995, ApJ, 452, 538
\re Gardner J.P., Cowie L.L., Wainscoat R.J. 1993, ApJ, 415, L9 
\re Glazebrook K., Peacock J.A., Miller L. Collins C.A. 1994, MNRAS, 
  266, 65
\re Harris H.C., Dahn C.C., Vrba F.J., Henden A.A., Lieber J., 
  Schmidt G.D. Reid I.N. 2000, ApJ, 524, 1000 
\re Hodgkin S.T., Oppenheimer B.R, Hambly N.C., Jameson R.F., 
  Smartt S.J., Steele I.A. 2000, Nature, 403, 57
\re Huchra J., Burg R. 1992, ApJ, 393, 90
\re Ibata R.A., Irwin M., Bienayme O., Scholz R., Guibert J. 2000, 
  ApJ, 532, L41
\re Ibata R.A., Richer H.B., Gilliland R.L., Scott D. 1999, ApJ, 
  524, L95
\re Jenkins C.R. Reid I.N. 1991, AJ, 101, 1595 
\re Kashikawa N, et al. 2000, in SPIE Conf. Vol. 4008, in press
\re Leggett S.K., Allard F., Hauschildt P.H. 1998, ApJ, 509, 836
\re Kirkpatrick J.D., Reid I.N., Liebert J., Cutri R.M., Nelson B., 
  Beichman C.A., Dahn C.C., Monet D.G., Gizis J.E., Skrutskie M.F. 
  1999, ApJ, 519, 802
\re McLeod B.A., Bernstein G.M., Rieke K.J., Tollestrup E.V., 
  Fazio G.G. 1995, ApJS, 96, 117 
\re Minezaki T., Kobayashi Y., Yoshii Y., Peterson B.A. 1998, ApJ,
  494, 111
\re Motohara K., Maihara T., Iwamuro F., Oya S., Imanishi M., 
  Terada H.,Goto M., Iwai J. et al.\ 1998, Proc.\ SPIE 3354, 659
\re Moustakas L.A., Davis M., Graham J.R., Peterson B.A., Yoshii Y. 
  1997, ApJ, 475, 445
\re Nakajima T., Oppenheimer B.R., Kulkani S.R., Golimowski D.A., 
  Matthews K., Durrance S.T. 1995, Nature, 378, 463
\re Nakajima T., Iwamuro F., Maihara T., Motohara K., Tsuji T., 
  Tamura M., Kashikawa N., Iye M. 2000, AJ, inpress 
\re Pozzetti L., Bruzual G., Zamorani G. 1996, MNRAS, 281, 953 
\re Pozzetti L., Madau P., Zamorani G., Ferguson H.C., Bruzual G. 
  1998, MNRAS, 298, 1133
\re Saracco P., D'Odorico S., Moorwood A., Buzzoni A., Cuby J.-G., 
  Lidman C. 1999, AA, 349, 751
\re Smail I., Hogg D.W., Yan L., Cohen J.G.  1995, ApJ, 449, L105
\re Teplits H.I., Gardner J.P., Malumuth E.M., Heap S.R. 1998, 
  ApJ, 507, L17
\re Thompson R.I., Storrie-Lombardi L.J., Weymann R.J., Rieke M.J., 
  Schneider G., Stobie E., Lytle D. 1999, AJ, 117, 17
\re Yan L., McCarthy P.J., Storrie-Lombardi L.J., Weymann R.J. 
  1998, ApJ, 503, L19
\re Totani T., Yoshii, Y.  2000, ApJ, 540, 81 
\re Yahata N., Lanzetta K.M., Chen H-W., Fernandez-Soto A., 
  Pascarelle M., Yahil A., Puetter R.C. 2000, ApJ, 538, 493
\re Yoshii Y. 1993, ApJ, 403, 552
\re Yoshii Y. Takahara F. 1988, ApJ, 326, 1 \\

\vspace*{10mm}
\clearpage

Figure Captions. \\


Figure 1.  The survey area of SDF overlaid on the Digitized Sky 
Survey (DSS) map. The near-infrared survey is performed in the 
$2^{\prime} \times 2^{\prime}$ region marked by a dashed square 
box.  The surrounding region, which we call flanking fields, is 
also shown enclosed by the solid line square box. \\


Figure 2.  $J$ band SDF image. \\


Figure 3.  $K'$ band SDF image. \\


Figure 4.  Two-color image composed from the $J$ and $K'$ band 
images. Image sizes are normalized to 0$.\hspace{-2pt}''$45 in FWHM. \\


Figure 5.  Multiple simulations of photometric measurements for 
mock objects embedded in survey images. An enlarged diagram with 
numerical fractions in the 0.5 mag cells is shown in the upper right 
corner to show how the measured total magnitude distributed against 
the model sources.  The lower panel shows diagrams 
of calculated completeness in both the $J$ (filled triangles) and 
the $K'$ band (filled circles). \\


Figure 6.  Number count vs. magnitude diagram of the $J$ band. 
The number counts corrected for completeness are plotted with filled 
symbols, while the raw counts are open symbols.  The circles, 
triangles, and squares denote total, isophotal, and aperture 
magnitudes, respectively. \\


Figure 7.  Same as Fig. 6, but for the $K'$ band. \\


Figure 8.  Galaxy number count vs. magnitude of the $J$ band. 
The data and authors of previous surveys are shown in the panel. \\


Figure 9.  Same as figure 8 but for the $K'$ band. \\


Figure 10.  AB magnitude plots of $J$, $H$, and $K'$ band data. Previous 
near-infrared surveys are also plotted for comparison. \\


Figure 12.  Integrated flux of SDF sources (filled circles) in each 
magnitude bin. Fluxes of other survey data are also shown. \\


Figure 12.  Color-magnitude relation for the SDF
sources. Objects detected in both the $J$ and $K'$ bands are plotted
by open circles. Objects detected either in the $J$ or $K'$ band
are represented by ``x'' or ``+'' marks, respectively. Thin solid 
lines are the color change vs. magnitude for representative
types of galaxy assuming a simple no-evolution model based on 
SED models of Yoshii \& Takahara (1988) and Coleman, Wu, \& Weedman 
(1980). Solid squares show the median color with standard deviation. 
Note that objects detected only in the $J$ or $K'$ band have 
been photometrically measured with same aperture sizes to 
determine $K'$ or $J$ band magnitudes, respectively,  as explained 
in the text. \\


Figure 13.  Examples of images of the four reddest objects found in the 
SDF.  $J$-band images are shown in the upper row, and $K'$-band images 
are in the lower row. \\


Figure 14.  FWHM vs.  $J - K'$ color diagram.  FWHM values are adopted 
from $K'$ band images.  Open stars represent objects identified as 
stars on the basis of the proclaimed criteria (see text).  Open 
circles are objects detected in both bands, while ``x'' and ``+'' 
marks represent objects selected only in the J-band and $K'$-band 
respectively. \\

\end{document}